\documentclass[epj,referee]{svjour}
\usepackage{graphics}
\begin{document}
\title{Frequency combinations in the magnetoresistance oscillations spectrum of a linear chain of
coupled orbits with a high scattering rate}

\author{ David Vignolles\inst{1}, Alain Audouard\inst{1}, Vladimir N. Laukhin\inst{2,3},
J\'{e}r\^{o}me B\'{e}ard\inst{1}, Enric Canadell\inst{3},
Nataliya G. Spitsina\inst{4}  \and Eduard B. Yagubskii\inst{4}
%
}                     
%
\mail{audouard@lncmp.org}
\institute{Laboratoire National des Champs Magn\'{e}tiques
Puls\'{e}s \thanks{UMR 5147: Unit\'{e} Mixte de Recherche CNRS -
Universit\'{e} Paul Sabatier - INSA de Toulouse}, 143 avenue de
Rangueil, 31400 Toulouse, France. \and Instituci\'{o} Catalana de
Recerca i Estudis Avan\c{c}ats (ICREA) 08210 Barcelona, Spain.
\and Institut de Ci\`{e}ncia de Materials de Barcelona, CSIC,
Campus de la UAB, 08193, Bellaterra, Spain \and Institute of
Problems of Chemical Physics, RAS, 142432 Chernogolovka, MD,
Russia}

\date{Received: \today / Revised version: date}
%
\abstract{The oscillatory magnetoresistance spectrum of the
organic metal (BEDO)$_5$Ni(CN)$_4\cdot$3C$_2$H$_4$(OH)$_2$ has
been studied up to 50 T, in the temperature range from 1.5 K to
4.2 K. In high magnetic field, its Fermi surface corresponds to a
linear chain of quasi-two-dimensional orbits coupled by magnetic
breakdown (MB). The scattering rate consistently deduced from the
data relevant to the basic $\alpha$ and the MB-induced $\beta$
orbits is very large which points to a significant reduction of
the chemical potential oscillation. Despite of this feature, the
oscillations spectrum exhibits many frequency combinations. Their
effective masses and (or) Dingle temperature are not in agreement
with either the predictions of the quantum interference model or
the semiclassical model of Falicov and Stachowiak.
} 
\authorrunning{D. Vignolles et al.}
\titlerunning{SdH oscillations spectrum of a linear chain of coupled orbits with a high scattering rate}
\maketitle
\section{Introduction}
\label{intro}

According to band structure calculations \cite{Ro04}, the Fermi
surface (FS) of many quasi-two-dimensional (q-2D) organic metals
with two charge carriers per unit cell originates from one
ellipsoidal hole tube whose area is therefore equal to that of the
first Brillouin zone (FBZ) area. In the extended zone scheme, this
tube can go over the FBZ boundaries along either one or two
directions leading to FS with different topologies. Due to gaps
opening at the crossing points, the resulting FS is composed of a
q-2D tube and a pair of q-1D sheets in the former case while, in
the second case, compensated electron and hole tubes are observed.
In high enough magnetic field, both of these FS's can be regarded
as networks of orbits coupled by magnetic breakdown (MB) which can
therefore be classified according to two types: the well known
linear chain of coupled orbits \cite{Pi62,Sh84} and the network of
compensated electron and hole orbits \cite{ClBr}, respectively.

In both cases, oscillatory magnetoresistance spectra exhibit
frequencies which are linear combinations of a few basic
frequencies. Magnetoresistance oscillations in linear chains of
coupled orbits have been studied, in particular in the case of the
compound $\kappa$-(BEDT-TTF)$_2$Cu(NCS)$_2$ \cite{Ha96,Ka96}. In
addition to the frequencies linked to the q-2D $\alpha$ orbit and
the MB-induced $\beta$ orbit (which corresponds to the hole tube
from which the FS is built) many combinations that can be
attributed to MB ($\beta$ + $\alpha$, $\beta$ + 2$\alpha$, etc.)
are observed in the Fourier spectra. Other Fourier components such
as $\beta$ - $\alpha$ \cite{Ha96} or $\beta$ - 2$\alpha$
\cite{Ka96} are also detected. These latter frequencies are
currently interpreted on the basis of quantum interference (QI),
although they are also observed in de Haas-van Alphen (dHvA)
oscillations spectra \cite{Me95,Uj97,St99}.

According to theoretical studies, both the field-dependent
modulation of the density of states due to MB \cite{Pi62,Sh84,LLB}
and the oscillation of the chemical potential \cite{mu} can also
induce frequency combinations in magnetoresistance and dHvA
oscillatory spectra. Nevertheless, the respective influence of
each of these contributions on the magnetoresistance and dHvA
spectra remains to be determined. Among them, the oscillation of
the chemical potential can be strongly reduced at high temperature
\cite{mu} and (or) for high scattering rate \cite{Ki03}.
Nevertheless, small effective masses are required in these cases
in order to observe quantum oscillations with a high enough
signal-to-noise ratio, at low enough magnetic field.

In this paper, we report on magnetoresistance experiments up to 50
T on (BEDO)$_5$Ni(CN)$_4\cdot$3C$_2$H$_4$(OH)$_2$ of which,
according to band structure calculations \cite{Du05}, the FS can
be regarded as a linear chain of coupled orbits. It is shown that
the crystals studied meet the above cited requirements, namely,
high scattering rates and low effective masses. Nevertheless, many
frequency combinations are observed in the oscillatory spectra.

\section{Experimental}
\label{sec:1}

The studied crystals, labelled  $\# 1$ and $\# 2$ in the
following, were synthesized by the electrocrystallization
technique reported in Ref. \cite{Du05}. They are roughly
parallelepiped-shaped with approximate dimensions (0.6 $\times$
0.3 $\times$ 0.1) mm$^3$, the largest faces being parallel to the
conducting \emph{ab} plane. Electrical contacts were made to the
crystals using annealed Pt wires of 20 $\mu$m in diameter glued
with graphite paste. Alternating current (10 to 100 $\mu$A, 20
kHz) was injected parallel to the \emph{c}* direction (interlayer
configuration). Magnetoresistance measurements were performed in
pulsed magnetic field of up to 50 T with pulse decay duration of
0.78 s, in the temperature range (1.5 - 4.2) K. A lock-in
amplifier with a time constant of 30 $\mu$s was used to detect the
signal across the potential leads.

\section{Results and discussion}

\begin{figure}                                                    
\centering \resizebox{\columnwidth}{!}{\includegraphics*{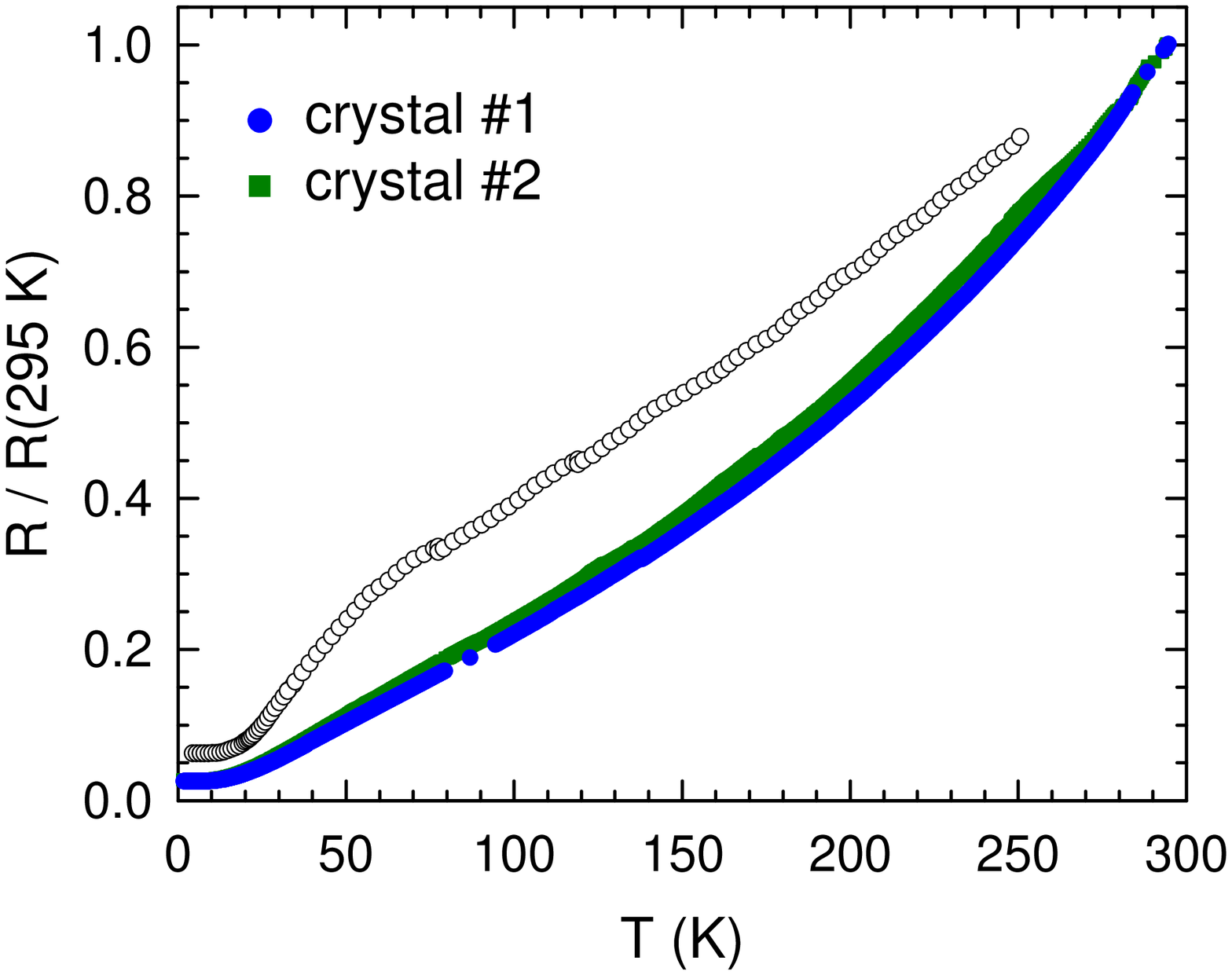}}
\caption{\label{RT} Temperature dependence of the normalized
interlayer resistance of crystals $\#$1 and $\#$2 (solid symbols).
Open symbols stand for the in-plane resistance data of Ref.
\cite{Du05}.}
\end{figure}

As reported in Figure \ref{RT}, the zero-field interlayer
resistance of the two studied crystals monotonically decreases as
the temperature decreases down to 1.5 K, as it is the case for the
in-plane resistance \cite{Du05}. The resistance ratio between room
temperature and 4.2 K (RR) is 39 and 36 for crystal $\#$1 and
$\#$2, respectively. These values (which are higher than for the
in-plane resistance data \cite{Du05} for which RR = 16) are much
lower than those reported for other compounds with similar FS such
as $\kappa$-(BEDT-TTF)$_2$Cu(NCS)$_2$ \cite{Ur88},
(BEDT-TTF)$_4$[Ni(dto)$_2$] \cite{Sc00} and
$\kappa$-(BEDT-TTF)$_2$I$_3$ \cite{We93}, namely, RR $\approx$
200, 400 and 8000, respectively.

\begin{figure}                                                    
\centering
\resizebox{\columnwidth}{!}{\includegraphics*{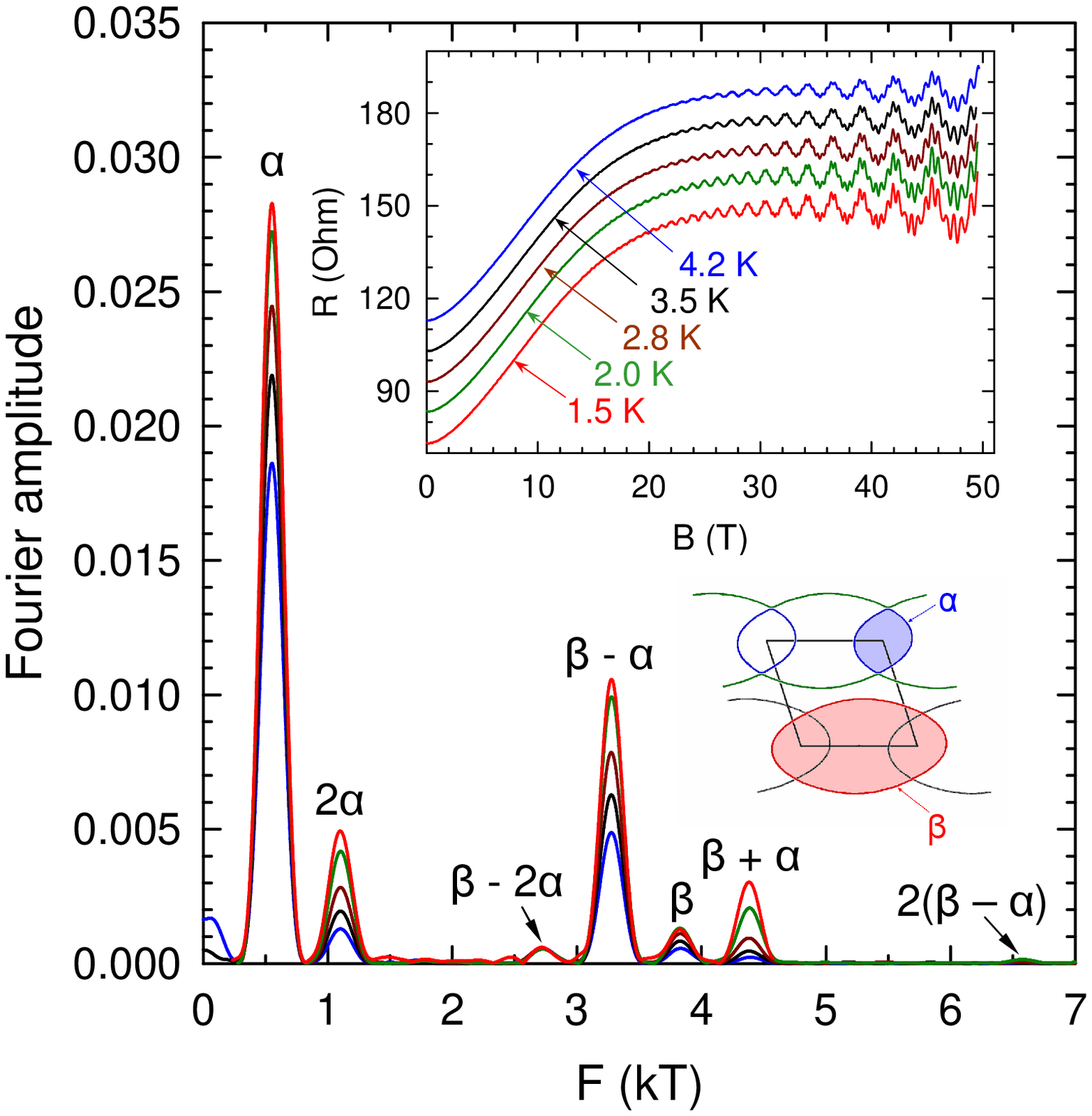}}
\caption{\label{R(B)_TF_SF} Fourier spectra deduced from the
magnetoresistance data displayed in the upper inset, in the field
range 32 T - 49.4 T. In this inset, the magnetoresistance curves
are shifted up by 10 ohms from each other for clarity. The lower
inset displays the Fermi surface deduced from band structure
calculations \cite{Du05}. }
\end{figure}

Magnetoresistance data of the two crystals yield consistent
results. In the following, we mainly focus on the data recorded
from crystal $\#$1 which was more extensively studied. As
generally observed for clean linear chains of coupled orbits, many
Fourier components are observed in the oscillatory
magnetoresistance spectra of which the frequencies are linear
combinations of those linked to the closed $\alpha$ and the
MB-induced $\beta$ orbits (see Figure \ref{R(B)_TF_SF} and Table
\ref{table}). The measured value of F$_{\beta}$ is in good
agreement with the expected value for an orbit area equal to that
of the FBZ. Indeed, according to crystallographic data at room
temperature \cite{Du05}, F$_{\beta}$ should be equal to 3839 T.
F$_{\alpha}$ corresponds to a cross section area of 14.4 percent
of the room temperature FBZ area. It is smaller than the value
predicted by band structure calculations \cite{Du05}, namely 20.8
percent of the FBZ area. Nevertheless, the obtained value is very
close to those reported for several compounds with similar FS such
as $\kappa$-(BEDT-TTF)$_2$Cu(NCS)$_2$ \cite{Ca94} and
$\kappa$-(BEDT-TTF)$_2$I$_3$ \cite{Ba95} for which the $\alpha$
orbit area amounts to 15 percent of the FBZ area. Band structure
calculations often lead to an overestimation of the area
associated with the $\alpha$ orbit for $\beta$"-BEDO salts with
this type of FS (see for instance ref. \cite{Ly02}). Because of
the scarcity of results it is not clear at this time if this
discrepancy is intrinsic to the theoretical approach used to
calculate the FS when considering BEDO donors, or due to an
unusual thermal contraction behavior of these salts leading to
larger changes than for similar sulfur-containing donors. However
this quantitative aspect is only of marginal importance for the
present study.
\\

\begin{table}                                                            
\caption{\label{table}Frequencies and effective masses (in m$_e$
units) linked to the various Fourier components (index i of Eqs.
\ref{LK} to \ref{RMB}) appearing in Figure \ref{R(B)_TF_SF}. }
\centering
\begin{tabular}{ccc}
\hline
i &F$_i$ (T)&m$^*_i$\\
\hline
$\alpha$&552 $\pm$ 3&1.15 $\pm$ 0.05\\
2$\alpha$&1104 $\pm$ 12&2.30 $\pm$ 0.08\\
$\beta$ - 2$\alpha$&2720 $\pm$ 40&0.3 $\pm$ 0.3\\
$\beta$ - $\alpha$&3278 $\pm$ 9&1.68 $\pm$ 0.05\\
$\beta$&3837 $\pm$ 24&1.80 $\pm$ 0.15\\
$\beta$ + $\alpha$&4388 $\pm$ 12&3.55 $\pm$ 0.15\\

\hline
\end{tabular}
\end{table}

With regard to the amplitude of the Fourier components appearing
in Figure \ref{R(B)_TF_SF}, the most striking feature is the small
amplitude of the $\beta$ component with respect to e.g. that
attributed to the MB-induced $\beta$ + $\alpha$ orbit. Indeed, in
the framework of the semiclassical model (see the discussion
below), the damping factors of the amplitude of  this latter
component are smaller than that related to $\beta$. The relatively
large amplitude of the $\beta$ - $\alpha$ component can also be
noticed. We have checked by changing the direction of the magnetic
field with respect to the crystalline axes that these features,
which are also observed in other organic metals with similar FS
\cite{Ha96,Ly02}, are not due to spin-zero phenomenon\footnote{A
change by a few degrees of the magnetic field direction with
respect to the normal to the conducting plane leads to a steep
(certainly due to the measured high Dingle temperature, see below)
decrease of both the $\beta$ - $\alpha$, $\beta$ and $\beta$ +
$\alpha$ Fourier components's amplitude.}.

\begin{figure}                                                        
\centering
\resizebox{\columnwidth}{!}{\includegraphics*{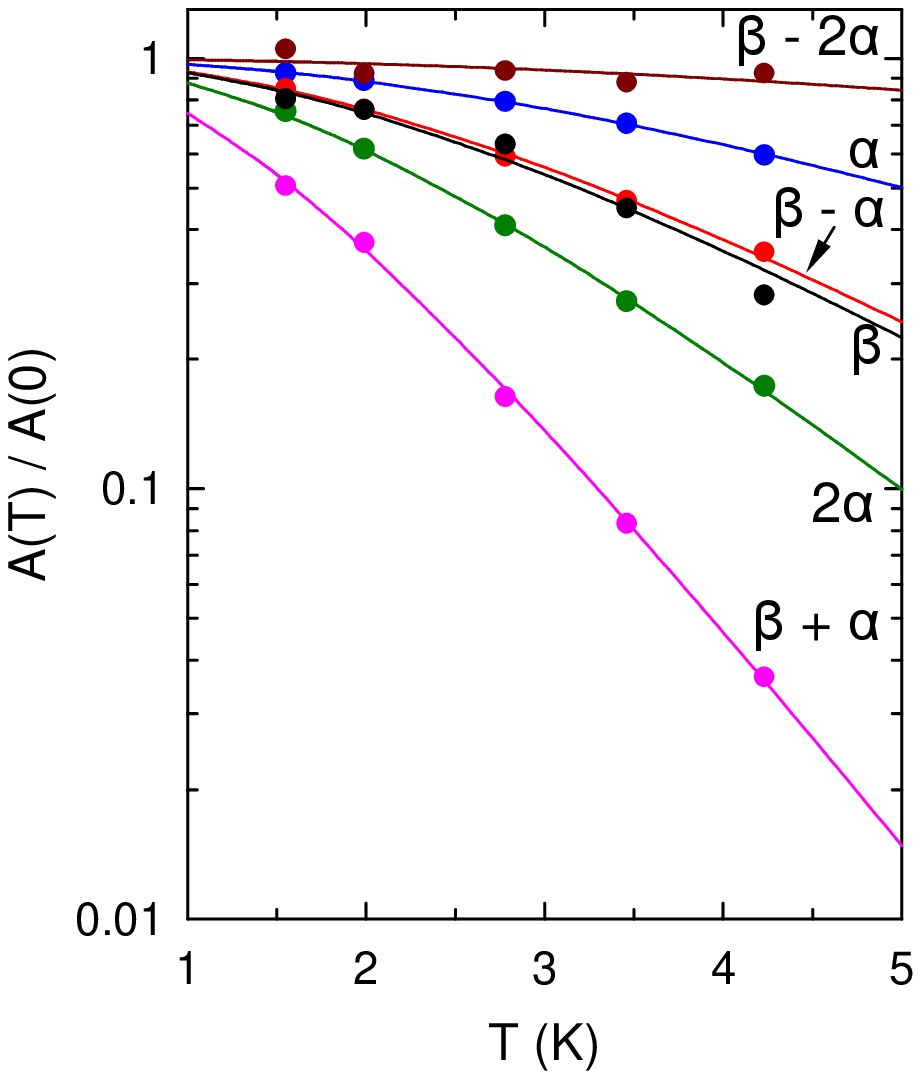}}
\caption{\label{mc} Temperature dependence of the amplitude of the
various Fourier components observed in Figure \ref{R(B)_TF_SF},
for a mean magnetic field value of 38 T. Solid lines are best fits
to Eq. \ref{LK}.}
\end{figure}

Let us consider in more detail the field and temperature
dependence of the amplitude of the Fourier components observed in
Figure \ref{R(B)_TF_SF}. In the framework of the Lifshits-Kosevich
(LK) model, the oscillatory magnetoresistance can be written as:

\begin{eqnarray}
\label{LK} \frac{R(B)}{R_{background}} = 1 +
\sum_{i}A_{i}cos[2{\pi}(\frac{F_i}{B}-\gamma_i)]
\end{eqnarray}

where R$_{background}$ is the monotonic part of the
field-dependent resistance R(B) and $\gamma_i$ is a phase factor.
The amplitude of the Fourier component with the frequency F$_i$ is
given by A$_{i} \propto$ R$_{Ti}$R$_{Di}$R$_{MBi}$R$_{Si}$, where
the spin damping factor (R$_{Si}$) depends only on the direction
of the magnetic field with respect to the conducting plane. The
thermal, Dingle and MB damping factors are respectively given by:

\begin{eqnarray}
\label{Rt}R_{Ti} =
\frac{{\alpha}T{}m{_i^*}}{Bsinh[{\alpha}T{}m{_i^*}/B ]}
\\R_{Di} =
exp[-{\alpha}T_Dm{_i^*}/B]\\
\label{RMB} R_{MBi} =
exp(-\frac{t_iB_{MB}}{2B})[1-exp(-\frac{B_{MB}}{B})]^{b_i/2}
\end{eqnarray}

where $\alpha$ = 2$\pi^2$m$_e$k$_B$/e$\hbar$ ($\simeq$ 14.69 T/K),
m$_i^*$ is the effective mass normalized to the free electron mass
(m$_e$), T$_D$ is the Dingle temperature and B$_{MB}$ is the MB
field. Integers $t_i$ and $b_i$ are respectively the number of
tunnelling and Bragg reflections encountered along the path of the
quasiparticle. Within this framework, high order harmonics of a
given component are regarded as frequency combinations (since e.g.
F$_{2\alpha}$ = 2F$_{\alpha}$), which is compatible with the
semiclassical model of Falicov and Stachowiak \cite{Sh84,Fa66}
(which predicts m$^*_{2\alpha}$ = 2m$^*_{\alpha}$, see below) and,
in any case, has no influence on the data analysis.

The effective mass linked to each of the observed Fourier
components, of which the frequencies are in the range from
F$_{\alpha}$ to F$_{\beta+\alpha}$, have been derived from the
temperature dependence of their amplitude for various field
windows. As an example, data are given in Figure \ref{mc} for a
mean magnetic field value of 38 T. It has been checked that the
values (listed in Table \ref{table}) determined for each of the
considered Fourier components remain field-independent within the
error bars in the range explored, in agreement with the
semiclassical model. It can be remarked that m$^*_{\alpha}$ and
m$^*_{\beta}$ are very low. Indeed, to our knowledge, the lowest
effective masses reported up to now for a linear chain of orbits
were observed in (BEDO-TTF)$_5$[CsHg(SCN)$_4$]$_2$ \cite{Ly02} for
which m$^*_{\alpha}$ = 1.6 $\pm$ 0.1 and m$^*_{\beta}$ = 3.0 $\pm$
0.3. In any case, the measured effective masses are much lower
than in the case of $\kappa$-(BEDT-TTF)$_2$Cu(NCS)$_2$ for which
m$^*_{\alpha}$ = 3.5 $\pm$ 0.1 and m$^*_{\beta}$ = 7.1 $\pm$ 0.5
\cite{Ha96}.\\

According to the coupled network model of Falicov and Stachowiak
\cite{Sh84,Fa66}, the effective mass value of a given MB orbit is
the sum of the effective masses linked to each of the FS pieces
constituting the considered orbit. Oppositely, in the framework of
the QI model \cite{St71}, the effective mass linked to a quantum
interferometer is given by the absolute value of the difference
between the effective masses linked to each of its two arms. This
leads, in the present case, to:

\begin{eqnarray}                                                             
\label{Eq:FalQI} m_{k\beta \pm l\alpha}^* = \mid  k \times
m_{\beta}^* \pm l \times m_{\alpha}^* \mid
\end{eqnarray}

for MB orbits or QI paths. Oppositely, analytical calculations of
the effective masses related to the frequency combinations induced
by the oscillation of the chemical potential in dHvA spectra of an
ideal two-band electronic system yield \cite{Fo05}:

\begin{eqnarray}                                                             
\label{Eq:Fo05}  m_{k\beta \pm l\alpha}^* = k \times m_{\beta}^* +
l \times m_{\alpha}^*
\end{eqnarray}

which is at variance with the above mentioned predictions for QI
paths. The failure of Eq. \ref{Eq:FalQI} in order to account for
the effective mass of the 2$^{nd}$ harmonics of $\alpha$ has been
reported for many 2D compounds \cite{Ha96b}. In contrast, Eq.
\ref{Eq:FalQI} is in full agreement with the data of 2$\alpha$
(see Table \ref{table}) as it has already been reported for the
2$\alpha$ \cite{Ha96,Ho96b} and 2$\beta$ \cite{To01} components
observed in magnetoresistance data of other linear chains of
orbits. However, according to the data of Table \ref{table},
neither Eq. \ref{Eq:FalQI}, nor Eq. \ref{Eq:Fo05} (which, strictly
speaking, is relevant only for dHvA spectra) account for the
effective mass of $\beta$ - $\alpha$. This result is at variance
with the magnetoresistance data of Refs. \cite{Ha96,Ly02,To01} for
which Eq. \ref{Eq:FalQI} holds\footnote{DHvA data of Refs.
\cite{Uj97,St99} yield large effective masses for $\beta$ -
$\alpha$, although lower than predicted by Eq. \ref{Eq:Fo05}.}.
Oppositely, Eq. \ref{Eq:FalQI} accounts for the effective mass of
$\beta$ - 2$\alpha$, as already observed for other compounds
\cite{Ka96,Ly02}. As for $\beta$ + $\alpha$, the effective mass
reported in  Table \ref{table} is slightly, although definitely,
higher than the value predicted by Eq. \ref{Eq:FalQI}.
Discrepancies between magnetoresistance data relevant to the MB
orbits and the predictions of the Falicov-Stachowiak model are
also reported for several other compounds with similar FS
\cite{Ha96}.
\\

\begin{figure} 
\centering
\resizebox{\columnwidth}{!}{\includegraphics*{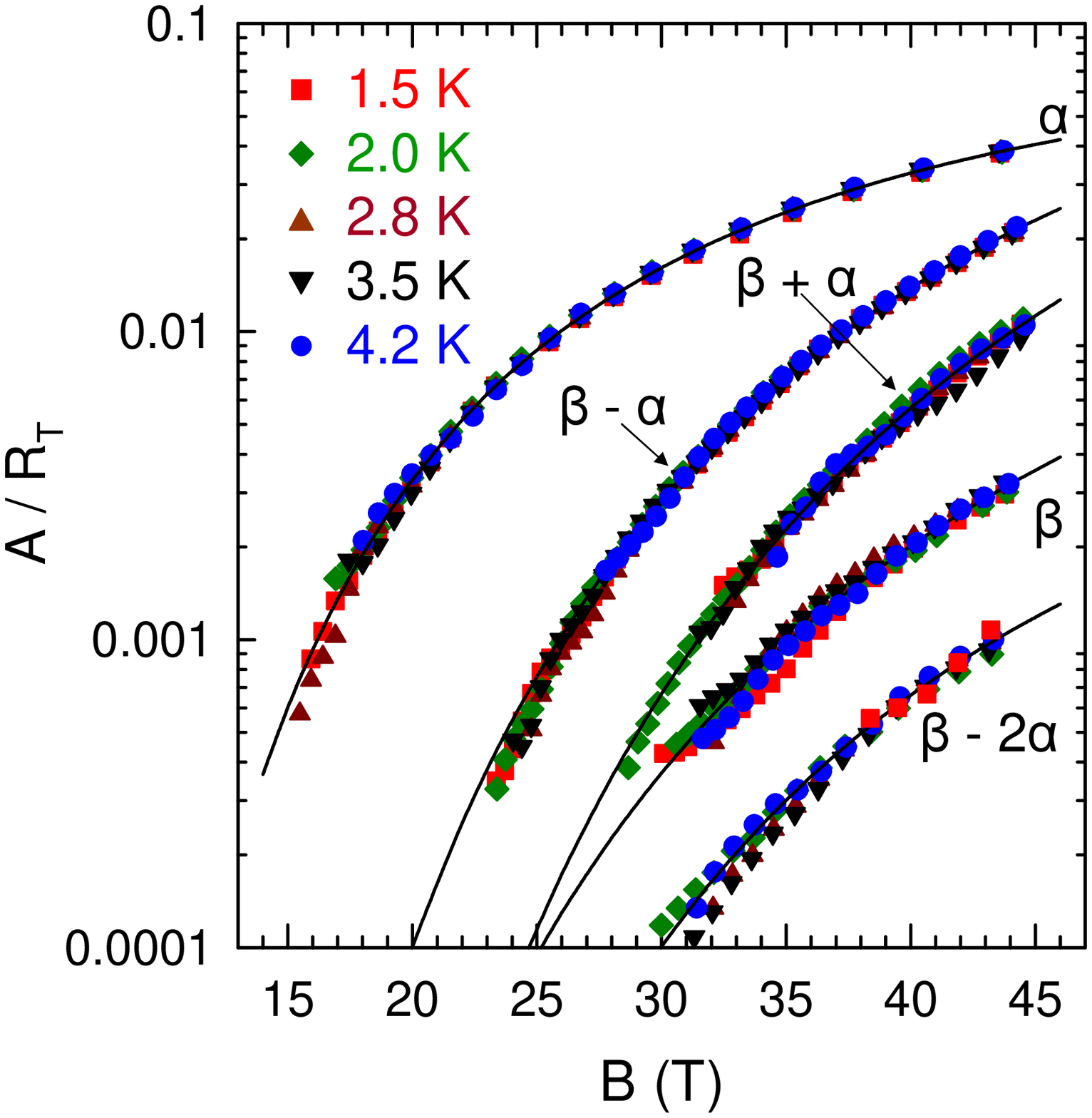}}
\caption{\label{Dingle} Field dependence of the
temperature-independent part of Fourier components deduced from
data in Figure \ref{R(B)_TF_SF}. A and R$_{T}$ are the amplitude
and the thermal damping factor of the considered Fourier
component, respectively. Solid lines are best fits of Eq. \ref{LK}
to the data.}
\end{figure}

These puzzling behaviours can be further checked by considering
the field dependence of the various component's amplitude.
According to Eqs. \ref{LK}-\ref{RMB}, their
temperature-independent part is given by:

\begin{eqnarray}                                                       
\label{Eq:AsurRT}\frac{A_i}{R_{Ti}} \propto
exp[-\frac{{\alpha}T_Dm{_i^*} + t_iB_{MB}/2}{B}] \nonumber\\
\times [1-exp(-\frac{B_{MB}}{B})]^{\frac{b_i}{2}}
\end{eqnarray}

The field dependence of this parameter is reported in Figure
\ref{Dingle} for most of the Fourier components considered in
Figure \ref{mc}. It can be remarked first that, as expected from
Eq. \ref{Eq:AsurRT}, the measured values of A$_i$ / R$_{Ti}$ are
actually temperature-independent in the range explored whatever
the considered Fourier component is. An estimation of T$_D$ can be
derived from the data of Fourier components for which t$_i$ = 0.
Assuming a low value for B$_{MB}$, namely B$_{MB}$ $\leq$ 2 T,
which is plausible owing to band structure calculations, data of
the $\alpha$ component yield T$_D$ = (7.0 $\pm$ 0.7) K [T$_D$ =
(7.9 $\pm$ 0.7) K for crystal $\#$2], which corresponds to a large
scattering rate ($\tau^{-1} \approx$ 6 $\times$ 10$^{12}$
s$^{-1}$). It must be noticed that the determined value of T$_D$
depend on B$_{MB}$ while the given error bars only account for the
experimental noise and the uncertainty on the relevant effective
mass. However, the B$_{MB}$ sensitivity of T$_D$ is moderate in
the considered case since T$_D$ = (6.7 $\pm$ 0.7) K and (6.3 $\pm$
0.6) K, assuming B$_{MB}$ = 10 T and 30 T, respectively. These
values are very close to each other and in any case equal within
the error bars. As for the MB field, the parameter A$_i$ /
R$_{Ti}$ goes to a maximum at B$_{max}$ = B$_{MB}$/ln[1 +
b$_i$B$_{MB}$/(2$\alpha$m$^*_i$T$_D$ + t$_i$B$_{MB}$)], according
to Eq. \ref{Eq:AsurRT}. B$_{max}$ should be reasonably less than
the maximum field reached in the experiments in order to derive a
reliable estimation of B$_{MB}$ \cite{ClBr}. In the present case,
the lowest value of B$_{max}$ is obtained in the case of the
$\alpha$ component. Unfortunately, the large Dingle temperatures
reported above lead to values of B$_{max}$ much larger than the
maximum field reached in the experiments, whatever the value of
B$_{MB}$ is. As a matter of fact, changing the value of B$_{MB}$
used in the fittings of Figure \ref{Dingle} (which, otherwise, are
obtained with B$_{MB}$ = 10 T) yields almost indiscernible curves
in the magnetic field range explored. According to Eq.
\ref{Eq:AsurRT}, B$_{MB}$ could also be derived from the data
linked to the $\beta$ component, assuming the Dingle temperature
relevant for this orbit is the same as for $\alpha$. However,
taking into account the above reported uncertainty on T$_D$,
B$_{MB}$ values compatible with the data relevant for the $\alpha$
component are in the range from 0 to 30 T which is probably valid
but constitutes a very large uncertainty. Besides, the
field-dependent data of Figure \ref{Dingle} for the components
$\beta$ - 2$\alpha$, $\beta$ - $\alpha$ and $\beta$ + $\alpha$
yield values of T$_D$ which are in poor agreement with the data
for $\alpha$ and $\beta$ \footnote{For B$_{MB}$ = 10 T, the Dingle
temperatures are (8.1 $\pm$ 0.7) K, (5.1 $\pm$ 0.6) K and dozens
of K, for $\beta$ - $\alpha$, $\beta$ + $\alpha$ and $\beta$ -
2$\alpha$, respectively.}. Remarkably, and at variance with the
preceding feature, data for the harmonics 2$\alpha$ yield T$_D$ =
(7.1 $\pm$ 0.7) K and T$_D$ = (8.2 $\pm$ 1.3) K for crystal $\#$1
and $\#$2, respectively which is in very good agreement with the
data for the $\alpha$ orbit.

Finally, the assumption of a negligibly small contribution of the
chemical potential oscillation to the oscillatory data can be
checked a posteriori. Indeed, according to \cite{Ki05}, the
cross-over temperature  T$_{co}$, above which the influence of the
chemical potential oscillation on the observed Fourier components
of a two-band electronic system becomes insignificant, is given by
T$_{co} \simeq$ 3B/$\alpha$m$^*$ - T$_D$. In the case of the
$\alpha$ orbit, which is the orbit with the lowest effective mass,
T$_{co}$ is equal to 1.5 K (i. e. the lowest temperature reached
in the experiments) at B = (48 $\pm$ 6) T, for crystal $\#$1. This
value lies very close to the upper boundary of the field range
considered in the experiments which indicates that, in addition to
QI, the only non-semiclassical phenomenon susceptible to
significantly contribute to the oscillatory spectra is the
coherent MB-induced modulation of the DOS \cite{Pi62,Sh84,LLB}.

\section{\label{sec:Conclusion}Summary and conclusion}

The FS of the organic metal
(BEDO)$_5$Ni(CN)$_4\cdot$3C$_2$H$_4$(OH)$_2$ corresponds to a
linear chain of orbits coupled by MB. Oscillatory
magnetoresistance spectra have been obtained in the temperature
range from 1.5 to 4.2 K in magnetic fields up to 50 T. The
temperature and field dependence of the Fourier components'
amplitude linked to the closed $\alpha$ and MB-induced $\beta$
orbits, analyzed in the framework of the semiclassical model of
Falicov and Stachowiak yields consistent results. In addition,
both the temperature and field dependencies of the component
linked to the 2$^{nd}$ harmonics of $\alpha$ are in good agreement
with the semiclassical model (same Dingle temperature as for
$\alpha$ and twice as large effective mass). The deduced
scattering rate $\tau^{-1}$ $\approx$ 6 $\times$ 10$^{12}$
s$^{-1}$ is very large that, according to \cite{Ki05}, allows for
considering that the contribution of the chemical potential
oscillation to the oscillatory magnetoresistance spectra is
negligible in the explored field and temperature ranges. Despite
of this feature, the recorded spectra exhibit frequency
combinations. Their effective masses and (or) Dingle temperature
are not in agreement with either the predictions of the QI model
or the semiclassical model of Falicov and Stachowiak. It can also
be remarked that their amplitude can be very large. In particular,
the amplitude of the components $\beta$ - $\alpha$ and $\beta$ +
$\alpha$ is larger than that of $\beta$. These results suggest
that additional contribution, such as the coherent MB-induced
modulation of the DOS \cite{Pi62,Sh84,LLB}, can play a significant
role in the observed frequency combinations.

\begin{acknowledgement}
 This work was supported by the French-Spanish exchange
 program between CNRS and CSIC (number 16 210), Euromagnet under the European Union contract
 R113-CT-2004-506239, DGI-Spain (Project BFM2003-03372-C03) and by Generalitat de Catalunya (Project 2005 SGR
 683). Helpful discussions with G. Rikken are acknowledged.
 \end{acknowledgement}

%

\begin{thebibliography}{}
%
%

\bibitem[1] {Ro04} R. Rousseau, M. Gener and E. Canadell, Adv.
Func. Mater. \textbf{14}, 201 (2004).

\bibitem [2] {Pi62} A. B. Pippard, Proc. Roy. Soc. (London)
\textbf{A270} 1 (1962).

\bibitem[3]{Sh84} D. Shoenberg, Magnetic Oscillations in Metals (Cambridge University Press, Cambridge,
1984).

\bibitem[4]{ClBr} D. Vignolles, A. Audouard, L. Brossard, S. I.
Pesotskii, R. B. Lyubovski\u{i}, M. Nardone, E. Haanappel and R.
N. Lyubovskaya, Eur. Phys. J. B, \textbf{31} 53 (2003); A.
Audouard, D. Vignolles, E. Haanappel, I. Sheikin,
R.~B.~Lyubovski\u{i} and R. N. Lyubovskaya, Europhys. Lett.
\textbf{71} 783 (2005).

\bibitem[5]{Ha96} N. Harrison, J. Caulfield, J. Singleton, P. H.
P. Reinders, F. Herlach, W. Hayes, M. Kurmoo and P. Day, J. Phys.
Condens. Matter \textbf{8} 5415 (1996).

\bibitem[6]{Ka96} M.V. Kartsovnik, G. Yu. Logvenov, T. Ishiguro,
W. Biberacher, H. Anzai, N.D. Kushch, Phys. Rev. Lett. \textbf{77}
2530 (1996).

\bibitem[7]{Me95} F. A. Meyer, E. Steep, W. Biberacher, P. Christ, A. Lerf, A.~G. M. Jansen,
 W. Joss, P. Wyder and K. Andres, Europhys. Lett. \textbf{32} 681
 (1995).

\bibitem [8]{Uj97}S. Uji, M. Chaparala, S. Hill, P. S.
Sandhu, J. Qualls, L.~Seger and J. S. Brooks, Synth. Met.
\textbf{85} 1573 (1997).

\bibitem [9]{St99} E. Steep, L. H. Nguyen, W. Biberacher,
H. M\"{u}ller, A.~G.~M.~Jansen and P. Wyder, Physica B
\textbf{259-261} 1079 (1999).

\bibitem[10]{LLB} P. S. Sandhu, J. H. Kim and J. S. Brooks, Phys. Rev. B
\textbf{56} (1997) 11566; J. Y. Fortin and T. Ziman, Phys. Rev.
Lett. \textbf{80} (1998) 3117; V. M. Gvozdikov, Yu V. Pershin, E.
Steep, A. G. M. Jansen and P. Wyder, Phys. Rev. B \textbf{65}
165102 (2002).

\bibitem[11]{mu} A. S. Alexandrov and A. M. Bratkovsky, Phys. Rev. Lett.
\textbf{76} (1996) 1308; Phys. Lett. A \textbf{234} 53 (1997) and
Phys. Rev. B \textbf{63} 033105 (2001); T. Champel, ibid.
\textbf{65} 153403 (2002); K. Kishigi and Y. Hasegawa, ibid.
\textbf{65} 205405 (2002).

\bibitem[12]{Ki03} K. Kishigi T. Ziman and H. Ohta, Physica B \textbf{329-333} 1156 (2003).

\bibitem[13]{Du05} A. D. Dubrovskii, N. G. Spitsina, L. I.
Buravov, G. V. Shilov, O. A. Dyachenko, E. B. Yagubskii, V. N.
Laukhin and E. Canadell, J. Mater. Chem. \textbf{15} (2005) 1248.

\bibitem [14]{Ur88} H. Urayama, H. Yamochi, G. Saito, K. Nozawa, T. Sugano, M. Kinoshita, S. Sato, K. Oshima,
A. Kawamoto and J. Tanaka, Chem. Lett. 55 (1988).

\bibitem[15]{Sc00} M.  Schiller, W. Schmidt, E. Balthes, D. Schweitzer, H. J. Koo, M. H. Whangbo, I. Heinen, T.
Klausa, P. Kircher, W. Strunz, Europhys. Lett. \textbf{51} 82.
(2000).

\bibitem [16]{We93} M. Weger, M. Tittelbach, E. Balthes, D.
Schweitzer and H. J. Keller, J. Phys.: Condens. Matter \textbf{5}
8569 (1993).

\bibitem[17]{Ca94} J. Caulfield, W. Lubczynski, F. L. Pratt, J.
Singleton, D. Y. K. Ko, W. Hayes, M. Kurmoo and P. Day, J. Phys.:
Condens. Matter  \textbf{6} (1994) 2911;  T. Biggs, A. K. Klehe,
J. Singleton, D. Bakker, J. Symington, P. Goddard, A. Ardavan, W.
Hayes, J. A. Schlueter, T. Sasaki and M. Kurmoo, ibid. \textbf{14}
L495 (2002).

\bibitem[18]{Ba95} E. Balthes, D. Schweitzer, I. Heinen, H.J.
Keller, W. Biberacher, A. G. M. Jansen, E. Steep, Synth. Metals
\textbf{70} 841 (1995); D. Schweitzer, E. Balthes, S. Kahlich, I.
Heinen, H.J. Keller, W. Strunz, W. Biberacher, A. G. M. Jansen, E.
Steep, ibid. \textbf{70} 857 (1995).

\bibitem[19]{Ly02} R. B. Lyubovski\u{i}, S. I. Pesotskii, M. Gener, R. Rousseau, E. Canadell, J. A. A. J. Perenboom,
V. I. Nizhankovskii, E. I. Zhilyaeva, O. A. Bogdanova  and R. N.
Lyubovskaya, J. Mater. Chem. \textbf{12} 483 (2002).

\bibitem[20]{Fa66} L. M. Falicov and H. Stachowiak, Phys. Rev. \textbf{147} 505
(1966).

\bibitem[21]{St71}R. W. Stark and C. B. Friedberg, Phys. Rev. Lett. \textbf{26} 556
(1971); R. W. Stark and C. B. Friedberg, J. of Low Temp. Phys.
\textbf{14} 111 (1974).

\bibitem[22]{Fo05} J. Y. Fortin, E. Perez and A. Audouard,
Phys. Rev. B \textbf{71} 155101 (2005).

\bibitem[23]{Ha96b} see N. Harrison, R. Bogaerts, P. H. P. Reinders,
J. Singleton, S. J. Blundell and F. Herlach, Phys. Rev. B
\textbf{54} 9977 (1996) and references therein.

\bibitem[24]{Ho96b} A. A. House, W. Lubczynski, S. J. Blundell,
J. Singleton, W. Hayes, M. Kurmoo and P. Day, J. Phys.:Condens.
Matter \textbf{8} 10377 (1996).

\bibitem[25]{To01} T. G. Togonidze, M. V. Kartsovnik, J. A. A. J. Perenboom, N. D. Kushch and H. Kobayashi,
Physica B \textbf{294-295} 435 (2001).

\bibitem[26]{Ki05} K. Kishigi and Y. Hasegawa, Synth. Met. \textbf{153} 381 (2005).

\end{thebibliography}
%

\end{document}